\documentclass[pra, aps, twocolumn, floatfix, showpacs]{revtex4}
\usepackage{graphicx, amsmath, amssymb, times}

\begin{document}
\title{Spin-orbit coupling induced Fulde-Ferrell-Larkin-Ovchinnikov-like Cooper pairing and skyrmion-like polarization textures in trapped optical lattices}
\author{M. Iskin}
\affiliation{
Department of Physics, Ko\c c University, Rumelifeneri Yolu, 34450 Sar{\i}yer, Istanbul, Turkey.
}
\date{\today}

\begin{abstract}

We study the interplay between the Zeeman field and spin-orbit coupling (SOC)
in harmonically trapped Fermi gases loaded into a two-dimensional single-band 
tight-binding optical lattice. Using the Bogoliubov-de Gennes theory, we find that 
the Zeeman field combined with a Rashba SOC gives rise to 
$(i)$ Fulde-Ferrell-like superfluidity and
$(ii)$ skyrmion-like polarization textures 
near the edges of the system. 
We also discussed the effects of interaction, temperature, SOC anisotropy and 
Zeeman field anisotropy on the superfluid ground state and polarization textures.

\end{abstract}
\pacs{05.30.Fk, 03.75.Ss, 03.75.Hh}
\maketitle

\section{Introduction}
\label{sec:intro}

The possibility of simulating non-Abelian artificial gauge fields with quantum 
Bose and Fermi gases in atomic systems has become one of the forefront 
research directions in the atomic and molecular physics 
community~\cite{nistsoc, chinasocb, chinasocf, mitsoc, engels, fu, williams, zhaireview, galitskireview}, 
primarily due to its direct connection to the topological phases of matter that have 
extensively been studied in the condensed-matter community 
in recent years~\cite{volovik, hasan, sczhang, wen}. 
In particular, the exciting possibility of the creation and observation of 
Majorana bound states in topological insulators, superconductors 
and superfluids is at the heart of topological quantum computation~\cite{tqc}. 
These quasiparticles can be created at the boundaries (edges) 
of non-Abelian topological phases, and they 
allow for non-local storage of quantum information that is protected 
from local perturbations by the bulk gap. Motivated by these theoretical
proposals, spin-orbit coupled Fermi gases have recently been created and 
detected near the quantum degeneracy limit by three groups~\cite{chinasocf, mitsoc, fu, williams}. 
While the Shanxi group in China studied the spin dynamics and
momentum distribution asymmetry in the equilibrium state as hallmarks of the
spin-orbit coupling~\cite{chinasocf, fu}, and the MIT group used a more direct 
approach and analyzed the energy-momentum dispersion, spin-orbit gap 
and spin composition of the quantum states~\cite{mitsoc}, the NIST group
has very recently identified a Feshbach resonance via its associated atomic 
loss feature~\cite{williams}. Thus, assuming that sufficiently low temperatures are 
experimentally attainable in the near future, the physics of Majorana 
bound states can be studied in the clean and controllable environment uniquely 
offered by the atomic systems~\cite{iskin, liu}.

Following the success of these initial 
experiments~\cite{nistsoc, chinasocb, chinasocf, mitsoc, engels, fu, williams} 
(see also the recent reviews~\cite{zhaireview, galitskireview}), there has 
been growing theoretical interest in studying the one-, two-, few-, 
and many-body properties of spin-orbit coupled Fermi gases. 
For instance, the stability and phase diagrams have 
been studied for finite and thermodynamic systems as functions 
of the interaction strength, SOC strength, population imbalance, Zeeman 
fields, SOC anisotropy, Zeeman field anisotropy, temperature, etc. in one, 
two and three dimensions (see e.g.~\cite{zheng, vjshenoy2b, dong2b, iskinZ, zhangyi, huZ, yiZ,puZ,seoZ}).
There also appeared some recent works on the normal-state properties of 
repulsive Fermi gases with short-range interactions in the upper branch 
of the spectrum~\cite{wmliu}, which share similarities with repulsive 
electron gases with long-range Coulomb interactions~\cite{ashrafi}.
The amount of knowledge gained in these recent works is overwhelming, 
and here we briefly quote the most recent ones that are concentrated on 
the possibility of creation and observation of 
Fulde-Ferrell-Larkin-Ovchinnikov (FFLO)~\cite{FF, LO} type 
spatially-modulated non-uniform superfluid phases under in-plane Zeeman 
field~\cite{zheng, vjshenoy2b, dong2b, iskinZ, zhangyi, huZ, yiZ,puZ,seoZ}. In sharp contrast 
to the out-of-plane Zeeman field, these works have shown that the 
non-uniform FFLO-like phases are energetically more favored than 
the uniform BCS-like phases in the case of an in-plane Zeeman field. 
It is important to note that the FFLO-type phases in spin-orbit coupled Fermi gases are 
stabilized mainly by the asymmetry of the Fermi surfaces in momentum space, 
and this mechanism is in contrast with that of the condensed-matter ones where 
they are stabilized by the symmetric Zeeman mismatch in momentum space.

Since all of these results are obtained through ansatz-based non-self-consistent 
momentum-space calculations~\cite{zheng, vjshenoy2b, dong2b, iskinZ, zhangyi, huZ, yiZ,puZ,seoZ}, 
one of our main objectives here is to investigate the stability of 
FFLO-like phases by solving the BdG equations in a 
self-consistent fashion. For this purpose, we study the interplay between 
the Zeeman field and SOC in two-dimensional Fermi gases~\cite{kohl, vogt, sommer} 
loaded into a single-band tight-binding optical lattice.
Our primary finding is that while the ground states of spin-orbit
coupled systems may have weak Fulde-Ferrell (FF)~\cite{FF} type non-uniform 
superfluid characters (i.e. phase modulations) but not a Larkin-Ovchinnikov 
(LO)~\cite{LO} one (i.e. amplitude modulations) under the 
out-of-plane Zeeman field, the FF character of the superfluids is stronger for the 
Rashba SOC under in-plane Zeeman field. The FF-type phase oscillations 
are most prominent along the direction that is perpendicular to the Zeeman field.
Therefore, our self-consistent real-space BdG results support recent findings 
on the thermodynamic continuum systems that are ansatz-based momentum-space 
calculations~\cite{zheng, vjshenoy2b, dong2b, iskinZ, zhangyi, huZ, yiZ,puZ,seoZ}.
We also comment on the effects of interaction, temperature, SOC anisotropy 
and Zeeman field anisotropy on the FFLO-like pairing and ground state of the system,
and note that since the superfluid order parameters modulate only towards the 
edges of the system, where the densities of fermions are low and the magnitudes 
of the order parameters are small, it may be difficult to detect these modulations 
in atomic systems at finite temperatures.

Furthermore, our secondary finding is that any non-zero combination of the 
Zeeman field and Rashba SOC induces not only an easy-axis polarization 
along the direction of the Zeeman field everywhere in the system but also 
a spatially-modulated (ring-shaped in magnitude) transverse polarization near the edges. 
This is in sharp contrast with the trapped systems with no-SOC 
(and also with the thermodynamic systems with SOC) where only an easy-axis 
polarization can be induced beyond a threshold Zeeman field. 
We show that the induced polarization textures are skyrmion-like~\cite{volovik, skyrme} 
finite-size effects, broadened by the trapping potential, 
and that their microscopic origin can be traced back to the counter-flow of 
spontaneous spin currents in the case of Rashba SOC. 
The skyrmion particles were originally proposed in late 1950s by the 
nuclear physicist T. Skyrme as a model for baryons~\cite{skyrme}, 
and they were first observed in condensed-matter physics with quantum-Hall 
ferromagnets as a result of the interplay between the Zeeman field and Coulomb 
interactions~\cite{sondhi, barrett}.
Note that similar skyrmion-like spin textures were previously predicted in atomic 
physics for rotating spinor BEC~\cite{khawaja, mueller} and spin-orbit coupled 
BEC~\cite{sinha, wu, rama}, where skyrmions are spontaneously produced 
by SOC in the latter case without rotation.
We also argue that the transverse polarization textures may be used to probe and 
characterize the topological phase transitions and the associated Majorana 
bound states in finite spin-orbit coupled Fermi gases, and comment on the 
effects of interaction, temperature, SOC anisotropy and Zeeman field 
anisotropy on the polarization textures.

The rest of this paper is organized as follows. In Sec.~\ref{sec:bdg}, first we 
introduce the mean-field Hamiltonian and then derive the self-consistency 
equations for the superfluid order parameter, total number of fermions, and 
out-of- and in-plane spin polarizations within the BdG framework. 
We numerically solve the resultant equations and discuss the obtained 
results in Sec.~\ref{sec:numerics}. Finally, the conclusions of this paper are 
briefly summarized in Sec.~\ref{sec:conclusions}.

\section{Bogoliubov-de Gennes theory}
\label{sec:bdg}

The results mentioned above are obtained within the self-consistent BdG 
theory in real-space as discussed next. First of all, we describe the spin-orbit coupled 
Fermi gases loaded into a two-dimensional single-band tight-binding optical lattice 
by the grand-canonical mean-field Hamiltonian,
\begin{align}
\label{eqn:ham}
H &= \sum_i \left(-t \sum_{\mathbf{\widehat{e}}} 
C_{i+\mathbf{\widehat{e}}}^\dagger \phi_{i+\mathbf{\widehat{e}},i} C_i 
+ \Delta_i c_{\uparrow i}^\dagger c_{\downarrow i}^\dagger + H.c.\right) \\
& - \sum_{i, \sigma} \left[ \left(\mu + s_\sigma h_z- V_i\right) c_{\sigma i}^\dagger c_{\sigma i} 
+ (h_x -i s_\sigma h_y) c_{\sigma i}^\dagger c_{-\sigma i} \right] \nonumber,
\end{align}
where the operator $c_{\sigma i}^\dagger$ ($c_{\sigma i}$) creates (annihilates) a
pseudo-spin $\sigma = \{ \uparrow, \downarrow\}$ fermion at lattice site $i$, the spinor 
$C_i^\dagger = (c_{\uparrow i}^\dagger, c_{\downarrow i}^\dagger)$ denotes the 
fermion operators collectively, 
$
\mathbf{\widehat{e}} = \{\mathbf{\widehat{x}}, \mathbf{\widehat{y}}\}
$ 
allows only nearest-neighbor hopping with amplitude $t$, and $H.c.$ is the Hermitian 
conjugate. For a generic non-Abelian gauge field $\mathbf{A}=(\alpha \sigma_y, -\beta\sigma_x)$, 
where $\sigma_e$ is the Pauli matrix and $\{\alpha, \beta\} \ge 0$ are independent 
parameters characterizing both the strength and the symmetry of the SOC, 
the $\uparrow$ and $\downarrow$ fermions gain
$
\phi_{i+\mathbf{\widehat{x}}, i} = e^{-i\alpha \sigma_y}
$
phase factors for hopping in the positive $\mathbf{\widehat{x}}$ direction and
$
\phi_{i+\mathbf{\widehat{y}}, i} = e^{i\beta \sigma_x}
$
phase factors for hopping in the positive $\mathbf{\widehat{y}}$ direction. 
In addition, the complex number 
$\Delta_i$ is the local mean-field superfluid order parameter (to be specified below), 
$\mu$ is the chemical potential, $s_\uparrow = - s_\downarrow = 1$, 
$\mathbf{h} \equiv (h_x, h_y, h_z)$ is the Zeeman field, and $V_i = V_0 r_i^2$ 
is the harmonic confining potential where the distance $r_i$ of site $i$ is measured 
from the center of the lattice. 

Using the Bogoliubov transformation, the mean-field Hamiltonian given in 
Eq.~(\ref{eqn:ham}) for a two-dimensional $L \times L$ square lattice can be 
compactly written as a $4L^2 \times 4L^2$ matrix-eigenvalue problem~\cite{edoko},
\begin{align}
\label{eqn:bdg.matrix}
\sum_{j} \left( \begin{array}{cccc}
T_{\uparrow \uparrow} & T_{\uparrow \downarrow} & 0 & \Delta \\
T_{\downarrow \uparrow} & T_{\downarrow \downarrow} & -\Delta & 0 \\ 
0 & -\Delta^* & -T_{\uparrow \uparrow}^* & -T_{\uparrow \downarrow}^* \\ 
\Delta^*& 0 & -T_{\downarrow \uparrow}^* & -T_{\downarrow \downarrow}^* 
\end{array} \right)_{ij}
&
\left( \begin{array}{c}
u_{nj}^\uparrow \\
u_{nj}^\downarrow \\
v_{nj}^\uparrow \\
v_{nj}^\downarrow
\end{array} \right) 
= \varepsilon_n
\left( \begin{array}{c}
u_{ni}^\uparrow \\
u_{ni}^\downarrow \\
v_{ni}^\uparrow \\
v_{ni}^\downarrow
\end{array} \right),
\end{align}
where $u_{ni}^\sigma$ and $v_{ni}^\sigma$ are the components of the $n$th quasiparticle
wave function at site $i$, and $\varepsilon_n \ge 0$ is the corresponding energy eigenvalue. 
Here, the offsite hopping and onsite energy terms are compactly written as 
\begin{align}
T_{\sigma \sigma'}^{ij} = - t_{\sigma \sigma'}^{ij} - 
& \left[ \left(\mu + s_\sigma h_z - V_i\right) \delta_{s_\sigma s_{\sigma'}} \right. \nonumber \\
& \left. + \left(h_x - i s_\sigma h_y\right) \delta_{s_\sigma,-s_{\sigma'}}
\right] \delta_{ij},
\end{align}
where $\delta_{ij}$ is the Kronecker delta. 
The non-vanishing nearest-neighbor hopping elements are 
$t_{\sigma\sigma}^{i,i+\mathbf{\widehat{x}}} = t \cos \alpha$ and 
$t_{\uparrow\downarrow}^{i,i+\mathbf{\widehat{x}}} = - t_{\downarrow\uparrow}^{i,i+\mathbf{\widehat{x}}} = -t \sin \alpha$
for the positive $\mathbf{\widehat{x}}$ direction, 
and $t_{\sigma\sigma}^{i,i+\mathbf{\widehat{y}}} = t \cos \beta$ and 
$t_{\uparrow\downarrow}^{i,i+\mathbf{\widehat{y}}} = t_{\downarrow\uparrow}^{i,i+\mathbf{\widehat{y}}} = i t \sin \beta$
for the positive $\mathbf{\widehat{y}}$ direction. 
Note that the hopping in the negative directions are simply the Hermitian 
conjugates, and also that the angles $\alpha$ and $\beta$ determine, respectively, 
the relative strength between the spin-conserving particle hopping and 
spin-flipping SOC terms in the $\mathbf{\widehat{x}}$ and $\mathbf{\widehat{y}}$ directions.

In this paper, we consider only the onsite interactions for which the off-diagonal 
couplings are $\Delta_{ij} = \Delta_i \delta_{ij}$ diagonal in the site index.
Therefore, Eq.~(\ref{eqn:bdg.matrix}) needs to be solved simultaneously with 
$
\Delta_i = g \langle c_{\uparrow i} c_{\downarrow i} \rangle,
$
where $g \ge 0$ is the strength of the onsite interaction between $\uparrow$ 
and $\downarrow$ fermions, and $\langle \cdots \rangle$ is a thermal average. 
In addition, we use $\mu$ to fix the total number of fermions $N = \sum_i n_i$,
where $0 \le n_i = \sum_\sigma \langle c_{\sigma i}^\dagger c_{\sigma i} \rangle \le 2$
gives the local fermion filling. Once the self-consistent solutions are obtained 
for the wave functions and the energy spectrum, it is a straightforward task to 
calculate any of the desired observables. 
For instance, we are interested in the local polarization vector 
$\mathbf{p_i} \equiv (p_{i x}, p_{i y}, p_{i z})$, the components of which follow from 
the expectation values of the Pauli spin matrices, 
i.e. $p_{i \nu} = \langle C_i^\dagger \sigma_\nu C_i \rangle$, and are given by
$p_{i x} = 2\textrm{Re} \langle c_{\uparrow i}^\dagger c_{\downarrow i} \rangle$,
$p_{i y} = 2\textrm{Im} \langle c_{\uparrow i}^\dagger c_{\downarrow i} \rangle$ and
$p_{i z} = \sum_\sigma s_\sigma \langle c_{\sigma i}^\dagger c_{\sigma i} \rangle$.
Thus, we need the following averages for our purposes
\begin{align}
\label{eqn:op}
\langle c_{\uparrow i} c_{\downarrow i} \rangle &= \sum_n \Big[ (v_{n i}^\uparrow)^* u_{n i}^ \downarrow f(\varepsilon_n)  + u_{n i}^\uparrow (v_{n i}^ \downarrow)^* f(-\varepsilon_n) \Big], \\
\label{eqn:inplane}
\langle c_{\uparrow i}^\dagger c_{\downarrow i} \rangle &= \sum_{n} \Big[ (u_{n i}^\uparrow)^* u_{n i}^\downarrow f(\varepsilon_n) + v_{n i}^\uparrow (v_{n i}^\downarrow)^* f(-\varepsilon_n) \Big], \\
\label{eqn:outofplane}
\langle c_{\sigma i}^\dagger c_{\sigma i} \rangle &= \sum_{n} \Big[ |u_{n i}^\sigma|^2 f(\varepsilon_n) + |v_{n i}^\sigma|^2 f(-\varepsilon_n) \Big],
\end{align}
where $f(x)=1/(e^{x/T}+1)$ is the Fermi-Dirac distribution function, $T$ is the
temperature and the Boltzmann constant $k_B$ is set to unity. We also define the
total polarization components as $P_\nu = \sum_i p_{i \nu}/N$, where $\nu \equiv\{x, y, z\}$.
Equations~(\ref{eqn:bdg.matrix})-(\ref{eqn:outofplane}) correspond to the generalization 
of the BdG equations to the case of spin-orbit coupled Fermi gases on optical lattices.

\section{Numerical results}
\label{sec:numerics}

Having established the BdG formalism, next we present our numerical solutions 
for the ground-state phases, which are performed on a $41a \times 41a$ square 
lattice with $N = 150$ fermions in total, where $a$ is the lattice spacing. 
We take $V_0 = 0.01t$ as the strength of the trapping potential, and discuss both 
the Rashba-type symmetric ($\alpha = \beta$) and asymmetric 
($\alpha \ne \beta$) SOC fields. Note that the experimentally more relevant equal 
Rashba-Dresselhaus (ERD) SOC~\cite{nistsoc, chinasocb, chinasocf, mitsoc,
engels,fu,williams,zhaireview, galitskireview} 
can be obtained by setting $\alpha = 0$. The effects of higher fermion numbers 
and finite temperature are also briefly mentioned towards the end of the paper. 

Before we discuss the aforementioned FFLO-like pairing and polarization textures, 
we make three important remarks. First, in the absence of a SOC, i.e. when $\alpha = \beta = 0$, 
we know that a sufficiently strong Zeeman field $\mathbf{h}$ (the threshold of which 
depends on $g$) can polarize the system along the easy-axis ($\mathbf{\widehat{h}}$) 
direction. Second, in the absence of a Zeeman field, i.e. when $\mathbf{h} \equiv (0,0,0)$,
we also know that the system is trivially unpolarized no matter what the SOC is. 
Third, while any combination of Zeeman field (no matter how weak the field is) 
and SOC in a thermodynamic system may produce a uniform polarization 
along the easy-axis direction, it does not induce any polarization in the 
transverse direction, i.e. perpendicular to $\mathbf{\widehat{h}}$. 
With these remarks in mind, next we show that any non-zero Zeeman field 
can induce intricate polarization textures near the edges of finite-size spin-orbit 
coupled Fermi gases under various circumstances.

\subsection{Out-of-plane Zeeman field}
\label{sec:out}

\begin{figure}[htb]
\centerline{\scalebox{0.82}{\includegraphics{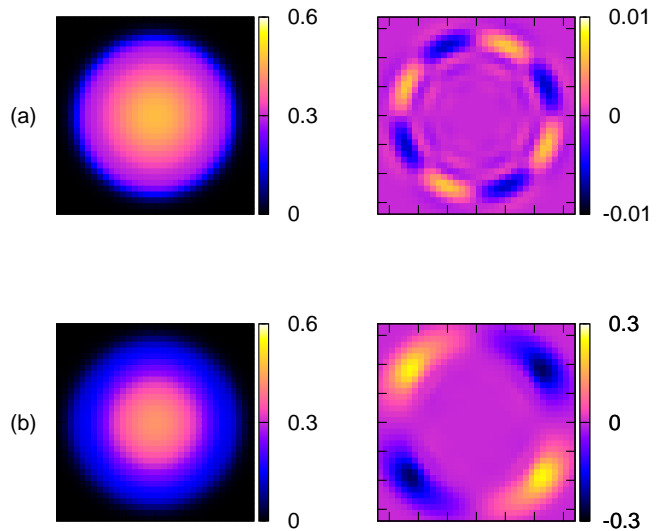}}}
\caption{\label{fig:outop} (Color online)
The color maps of the amplitudes ($|\Delta_i|$ - left column) and phases 
($\phi_i$ - right column) of the order parameters $\Delta_i = |\Delta_i| e^{i\phi_i}$ 
are shown for an out-of-plane Zeeman field $\mathbf{h} \equiv (0, 0, 0.5t)$ 
at $g = 3t$ and $T = 0$, where $\alpha = \beta = \pi/4$ in (a), 
and $\alpha = \pi/40$ and $\beta = \pi/4$ in (b). 
}
\end{figure}

Let us first consider an out-of-plane Zeeman field $\mathbf{h} \equiv (0, 0, h_z \ne 0)$,
which is perpendicular to our square lattice. The magnitudes and phases of typical 
ground-state order parameters are illustrated for the Rashba and ERD-like 
SOCs in Figs.~\ref{fig:outop}(a) and~\ref{fig:outop}(b), respectively, for $h_z = 0.5t$.
The phases of the order parameters clearly show the $C_4$ and $C_2$ symmetries
of the Hamiltonian for the Rashba and ERD-like SOCs, respectively.
These figures suggest that the ERD-like SOC has a stronger 
FF-type non-uniform superfluid character where the phase of the order 
parameter has a much larger spatial modulation. Note that the phases 
have both angular and radial oscillations towards the edges of the system 
where the densities of the fermions are low and the magnitudes of the order 
parameters are small. Therefore, it may be difficult to detect these modulations 
in atomic systems at finite $T$. However, since the spatial profiles of the 
magnitudes of the order parameters do not have any zeros 
(nodes), our results do not feature any LO-type non-uniform superfluidity. 
We emphasize that these effects become weaker and weaker with decreasing 
$h_z$ in such a way that all of the local phases of the order parameters 
vanish as $h_z \to 0$.

\begin{figure}[htb]
\centerline{\scalebox{0.75}{\includegraphics{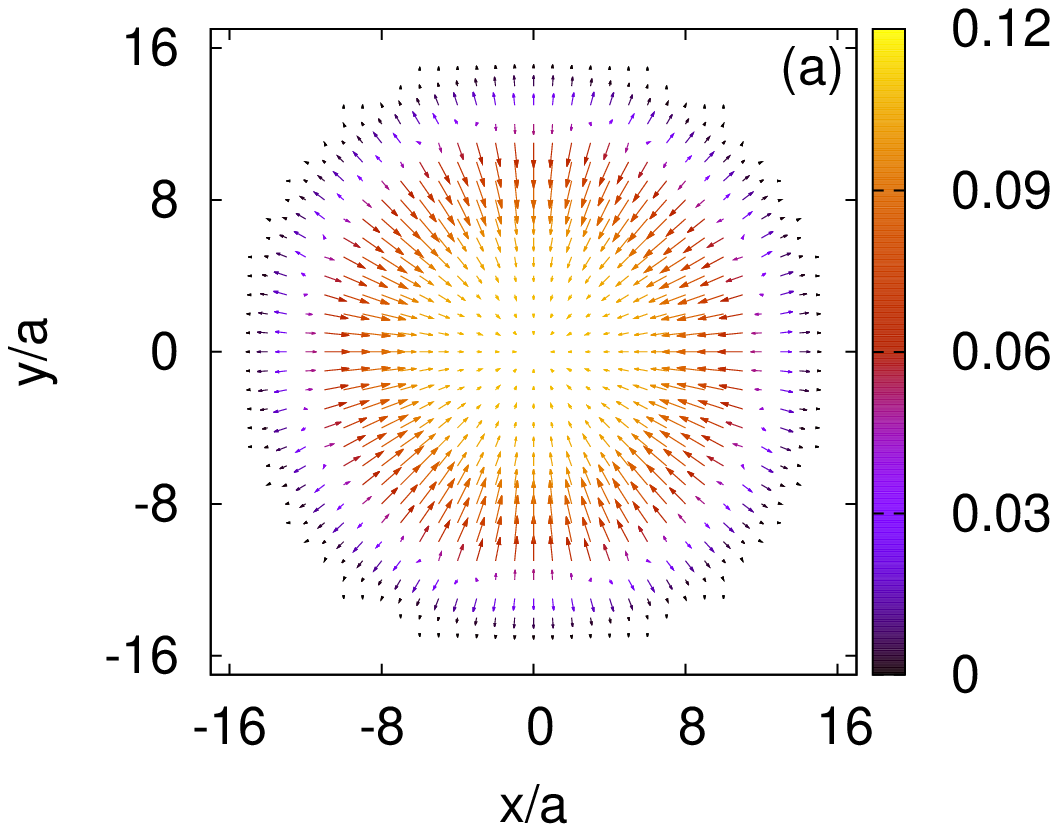}}}
\centerline{\scalebox{0.75}{\includegraphics{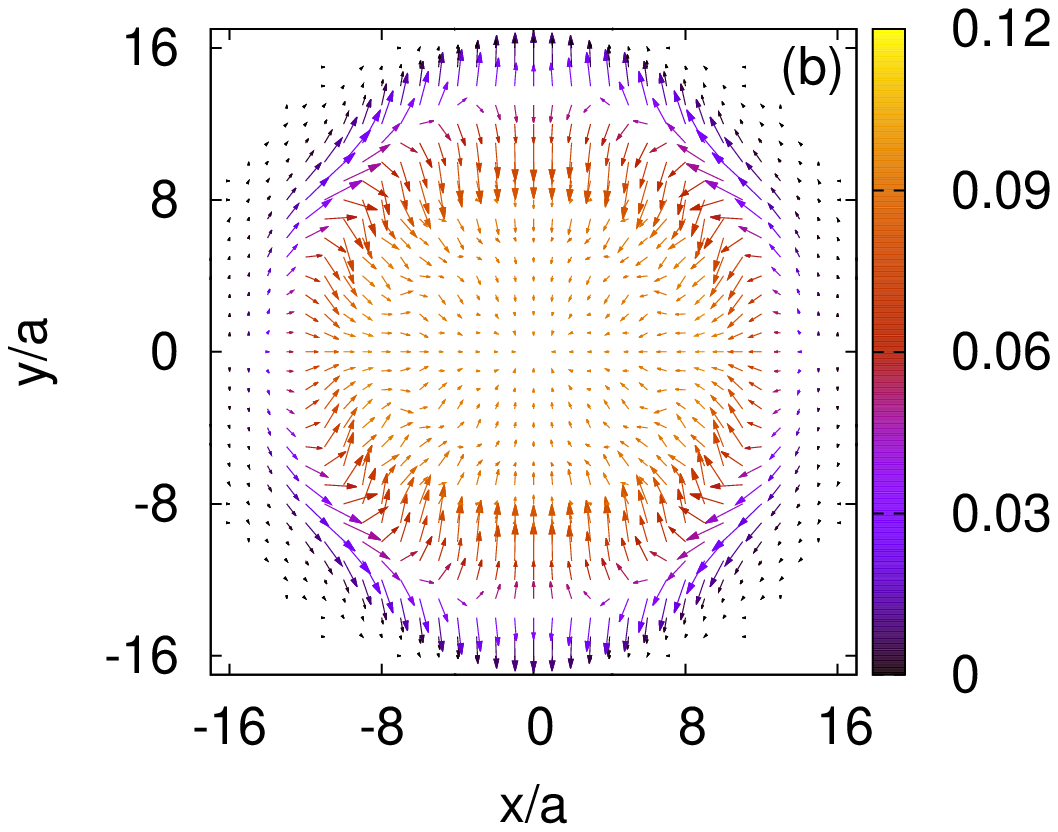}}}
\caption{\label{fig:out} (Color online)
Two-dimensional vector maps of the transverse polarizations $(-p_{i x}, -p_{i y})$ 
are shown for an out-of-plane Zeeman field $\mathbf{h} \equiv (0, 0, 0.5t)$ 
at $g = 3t$ and $T = 0$. Here, $\alpha = \beta = \pi/4$ in (a), and $\alpha = \pi/40$ and 
$\beta = \pi/4$ in (b). The largest arrows correspond approximately to 0.01, 
and the easy-axis polarizations $p_{i z}$ are illustrated with color maps. 
}
\end{figure}

The corresponding ground-state polarization textures are illustrated in 
Fig.~\ref{fig:out}, where we show two-dimensional vector maps of the transverse 
polarizations $(-p_{i x}, -p_{i y})$ together with color maps of the easy-axis 
polarizations $p_{i z}$. Here, we set $g = 3t$ but emphasize that
setting it to 0 does not lead to any significant change in the results. 
These figures again clearly show the $C_4$ and $C_2$ symmetries of the 
Hamiltonian for the Rashba and ERD-like SOCs, respectively.
First of all, the not-so-interesting $p_{i z}$ is finite everywhere in the trap with
its maximum value at the center of the system in both figures, and it 
gradually decreases to zero towards the edges. 
The case of Rashba SOC is shown in Fig.~\ref{fig:out}(a), where we find 
that $p_{i x} \ne 0$ and $p_{i y} \ne 0$ in general, except for the center of the trap.
In the ERD-like case when $\alpha \to 0$ but $\beta = \pi/4$, we see in Fig.~\ref{fig:out}(b) 
that while $p_{i x} \to 0$ everywhere in the system, $p_{i y}$ remains mostly unchanged. 
Therefore, in the ERD case when $\alpha = 0$, a domain-wall is formed on the 
$x$ axis where $p_{i x} = p_{i y} = 0$, and such a limiting behavior can be extracted 
from Fig.~\ref{fig:out}(b). Similarly, when $\beta = 0$ but $\alpha \ne 0$, 
a domain-wall forms on the $y$ axis where $p_{i x} = p_{i y} = 0$ (not shown).
Thus, we conclude that a non-zero out-of-plane Zeeman field no matter how
small it is (not shown) induces spatially modulated transverse polarizations 
only along those directions where there is SOC. In most cases, the ratio
of the transverse to the easy-axis polarizations are around $\%5-\%10$.
However, we emphasize that while the total easy-axis polarizations are 
$P_z \approx 0.3$ and $P_z \approx 0.36$ in Figs.~\ref{fig:out}(a)
and~\ref{fig:out}(b), respectively, the total transverse polarizations 
vanish, i.e. $P_x = 0 = P_y$, as one may expect.

Note that the transverse polarizations change sign in space, and that their 
magnitudes $\sqrt{p_{i x}^2 + p_{i y}^2}$ show ring-shaped 
structures. In Fig.~\ref{fig:out}(a), in addition to the broad ring ranging mostly 
between $7a \lesssim r_i \lesssim 12a$, there is also a narrow one with a very
weak peak around $r_i \approx 14a$. Similarly, in Fig.~\ref{fig:out}(b), there are 
two incomplete rings especially along the $y$ axis, and they both have comparable 
peaks around $r_i \approx \{10.5a, 14.5a\}$. We also find that increasing 
the size of the square lattice pushes the ring-shaped structures further away 
from the center, and reducing the strength of the trapping potential intensifies 
them around a narrower region near the edges, eventually both leaving 
no transverse polarization near the center. 
In the box-potential limit when $V_0 \to 0$, we find that while the continuum- and 
edge-state contributions to the transverse polarizations are competing 
with each other for low $h_z$ values, the latter contribution gets stronger 
with increasing $h_z$, and eventually dominates beyond the $h_z$ threshold 
for the creation of zero-energy (Majorana) edge-bound states.
These findings suggest that the transverse polarizations observed here are 
finite-size edge effects, broadened by the trapping potential, and also that
the working mechanism is similar to the one that is responsible for the creation 
of edge-bound Majorana states.
We note that similar spin textures are referred to as skyrmions in the contexts 
of rotating spinor BEC~\cite{khawaja, mueller} and spin-orbit coupled 
BEC without rotation~\cite{sinha, wu, rama}. In particular, we especially note the great 
similarity between Fig.~\ref{fig:out}(a) presented here and Fig. 5(a) of Ref~\cite{rama}. 

To understand the microscopic origin of these textures, next we employ the 
local-density approximation and analyze the single-particle excitation spectrum 
of the local system. The spectrum of the local Hamiltonian in momentum space 
involves two quasihole and two quasiparticle branches that are given by
\begin{align}
\label{eqn:ek}
E_{\mathbf{k} i, \pm}^2 &= \xi_{\mathbf{k} i}^2 + h_z^2 + |s_{\mathbf{k}}|^2 + |\Delta_i|^2 \nonumber \\
& \pm 2\sqrt{h_z^2(\xi_{\mathbf{k},i}^2 + |\Delta_i|^2) + \xi_{\mathbf{k} i}^2 |s_{\mathbf{k}}|^2},
\end{align}
where
$
\xi_{\mathbf{k} i} = -2t[\cos\alpha \cos(k_x a) + \cos\beta \cos(k_y a)] - \mu_i
$
is the shifted kinetic energy and
$
|s_{\mathbf{k}}|^2 = 4t^2[\sin^2\alpha \sin^2(k_x a) + \sin^2\beta \sin^2(k_y a)]
$
is the SOC contribution. Here, the local chemical potential $\mu_ i = \mu - V_i$ 
includes the trapping potential. We immediately see that the minus 
branches can become gapless at some $\mathbf{k}$-space points, 
i.e. $E_{\mathbf{k_0} i, -}^2 = 0$, and therefore the location of zero-energy 
states are determined by the following conditions: 
($i$) $|s_{\mathbf{k_0}}| = 0$ and 
($ii$) $h_z = \sqrt{\xi_{\mathbf{k_0} i}^2 + |\Delta_i|^2}$. 
While the Rashba SOC satisfies the former condition at four points 
$\mathbf{k_0} \equiv \{(0, 0)$; $(0, \pi)$; $(\pi, 0)$; $(\pi, \pi)$\}, the ERD SOC 
satisfies it only at two points $\mathbf{k_0} \equiv \{(k_x, 0)$; $(k_x, \pi)$\}. 

It is clear that the latter condition ($ii$) is easier to satisfy towards the edges 
of the system when $|\Delta_i| \to 0$, and since the transverse polarizations 
are found to be very similar for $g = 3t$ and $g = 0$, we may set $|\Delta_i| = 0$ 
for our purpose. When this is the case, the condition ($ii$) becomes
$
\mu_i = \pm h_z \pm 2t(\cos \alpha \pm \cos \beta)
$
for the Rashba SOC, and
$
\mu_i = \pm h_z - 2t[\cos(k_x a) \pm \cos \beta]
$
for the ERD SOC, where all $\pm$ combinations are possible and $|\cos(k_x a)| \le 1$. 
For the parameters of Figs.~\ref{fig:out}(a) and~\ref{fig:out}(b), 
where $\mu \approx -1.76 t$, these conditions are satisfied at two distances
$
r_i \approx \{ 7.5a, 12.5a \}
$
and
$
r_i \approx \{ 10.7a, 14.6a \},
$
respectively, which are very close to our numerical results given above. 
Thus, we conclude that the microscopic origin of the transverse polarizations 
can be traced back to the changes in the momentum-space topology of the 
single-particle excitation spectrum of the local system.

In the case of Rashba SOC, we can also interpret these textures as a direct 
consequence of counter-flow of spontaneously-induced spin currents~\cite{edoko}. 
This is because the Rashba SOC gives rise to an effective momentum-dependent 
in-plane magnetic field in the direction that is perpendicular to the in-plane 
momentum. Since the induced spin currents are circulating along the trap 
edges, i.e. the in-plane momentum is in the azimuthal direction, the induced 
in-plane spin texture is in the radial direction. The relative contribution between
the radially-outward and -inward helicity bands depends on the local chemical 
potential $\mu_i$, and this competition produces the spatial structure of the
spin textures such as the one illustrated in Fig.~\ref{fig:out}(a). Note that the 
time-reversal symmetry of the spins must be broken via e.g. the Zeeman field 
in order to have a non-zero polarization in any particular direction.

\subsection{In-plane Zeeman field}
\label{sec:in}

\begin{figure}[htb]
\centerline{\scalebox{0.82}{\includegraphics{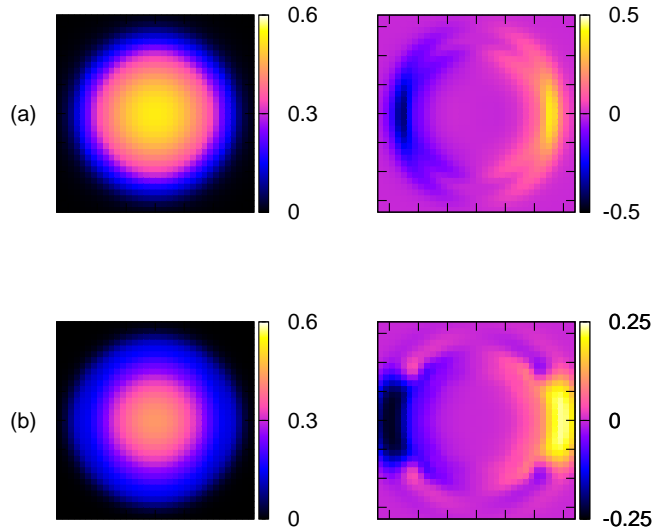}}}
\caption{\label{fig:inop} (Color online)
The color maps of the amplitudes ($|\Delta_i|$ - left column) and phases 
($\phi_i$ - right column) of the order parameters $\Delta_i = |\Delta_i| e^{i\phi_i}$ 
are shown for an in-plane Zeeman field $\mathbf{h} \equiv (0, 0.5t, 0)$ 
at $g = 3t$ and $T = 0$, where $\alpha = \beta = \pi/4$ in (a), 
and $\alpha = \pi/40$ and $\beta = \pi/4$ in (b). 
}
\end{figure}

Next, we consider an in-plane Zeeman field $\mathbf{h} \equiv (0, h_y \ne 0, 0)$,
which lies in the $\mathbf{\widehat{y}}$ direction parallel to our square lattice.
The magnitudes and phases of typical ground-state order parameters are 
illustrated for the Rashba and ERD-like SOCs in Figs.~\ref{fig:inop}(a) 
and~\ref{fig:inop}(b), respectively, for $h_y = 0.5t$.
In sharp contrast to the out-of-plane Zeeman case discussed above, this 
comparison clearly shows that the Rashba SOC has a much stronger 
FF-type non-uniform superfluid character in the in-plane Zeeman case, 
without again featuring any LO-type order parameter node. Note again 
that the phases have both angular and radial oscillations towards the edges 
of the system, and the FF-type oscillations are most prominent along the 
$x$ direction, i.e. perpendicular to the direction of the Zeeman field.
We emphasize that these effects become weaker and weaker with decreasing 
$h_y$ in such a way that all of the local phases vanish as $h_y \to 0$.

\begin{figure}[htb]
\centerline{\scalebox{0.75}{\includegraphics{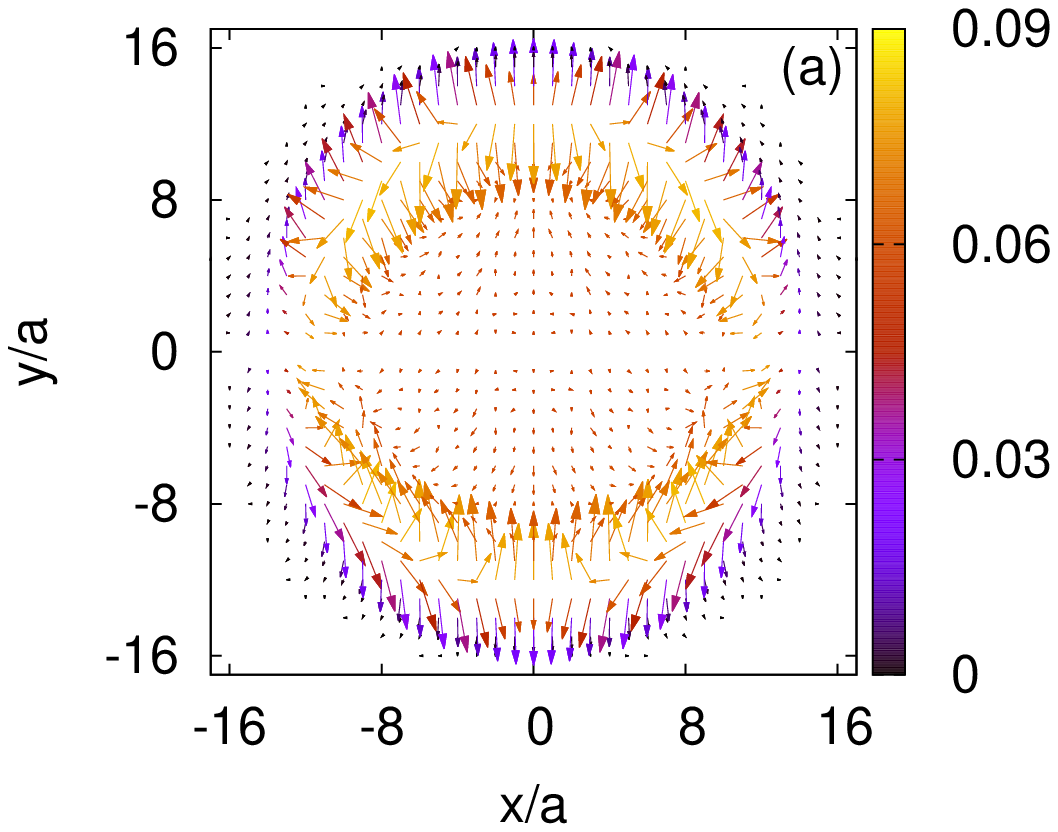}}}
\centerline{\scalebox{0.75}{\includegraphics{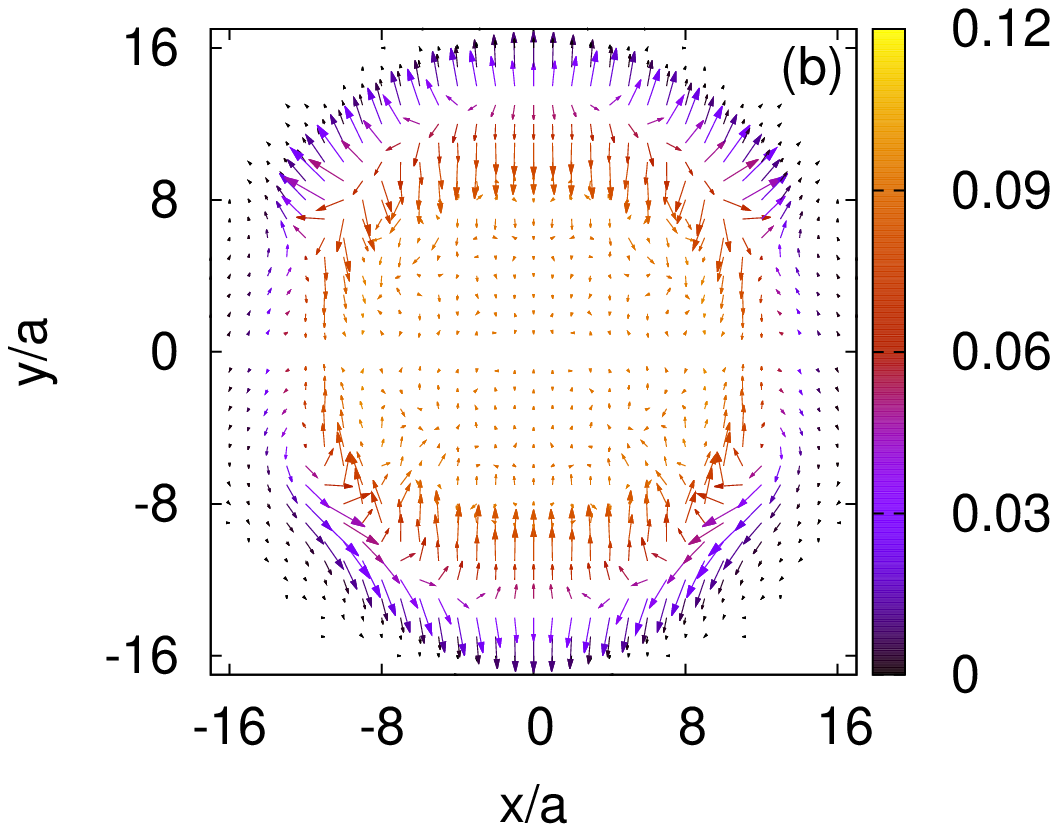}}}
\caption{\label{fig:in} (Color online)
Two-dimensional vector maps of the transverse polarizations $(p_{i x}, p_{i z})$ 
are shown for an in-plane Zeeman field $\mathbf{h} \equiv (0, 0.5t, 0)$ 
at $g = 3t$. Here, $\alpha = \beta = \pi/4$ in (a), and $\alpha = \pi/40$ and 
$\beta = \pi/4$ in (b). The largest arrows correspond approximately to 0.014, 
and the easy-axis polarizations $p_{i y}$ are illustrated with color maps. 
}
\end{figure}

The corresponding ground-state polarization textures are illustrated in Fig.~\ref{fig:in}, 
where we show two-dimensional vector maps of the transverse polarizations
$(p_{i x}, p_{i z})$ together with color maps of the easy-axis polarizations 
$p_{i y}$. Here, we set $g = 3t$ but setting $g$ to 0 again 
does not lead to any significant change in the results. 
First of all, the not-so-interesting $p_{i y}$ is finite everywhere in the trap with
its maximum value near the center in both figures, and it gradually decreases 
to zero towards the edges. The case of Rashba SOC is shown 
in Fig.~\ref{fig:in}(a), where we find that $p_{i x} \ne 0$ and 
$p_{i z} \ne 0$ in general, except for a domain-wall on the $x$ axis 
where $p_{i x} = p_{i z} = 0$. In the ERD-like SOC when $\alpha \to 0$ 
but $\beta \ne 0$, we see in Fig.~\ref{fig:in}(b) that while $p_{i x} \to 0$ 
everywhere in the system, $p_{i z}$ remains mostly unchanged. 
Therefore, similar to the Rashba case, the ERD case also has a domain-wall 
that is formed on the $x$ axis where $p_{i x} = p_{i z} = 0$,
and such a limiting behavior can be extracted from Fig.~\ref{fig:in}(b).
On the other hand, when $\beta = 0$ but $\alpha \ne 0$, there is not any 
transverse polarization in the entire system, i.e. $p_{i x} = p_{i z} = 0$ 
for every $i$ (not shown). Thus, we conclude that, when a non-zero in-plane Zeeman 
field (no matter how small it is) is not perpendicular to the direction of the 
SOC, a spatially modulated polarization is induced in the transverse direction. 
In most cases, the ratio of the transverse to the easy-axis polarizations 
are around $\%5-\%15$.
However, we emphasize that while the total easy-axis polarizations are 
$P_y \approx 0.37$ and $P_y \approx 0.36$ in Figs.~\ref{fig:in}(a) and~\ref{fig:in}(b), 
respectively, the total transverse polarizations again vanish, i.e. $P_x = 0 = P_z$.

Similar to the out-of-plane Zeeman case, we note that the magnitudes 
of the transverse polarizations $\sqrt{p_{i x}^2 + p_{i z}^2}$ show ring-shaped 
structures in both Figs.~\ref{fig:in}(a) and~\ref{fig:in}(b), where the peaks occur, 
respectively, at $r_i \approx \{10.5a, 13.5a\}$ and $r_i \approx \{10.5a, 14.5a\}$ 
away from the center of the trap especially along the $y$ axis. 
The microscopic origin of these structures can again be traced back to the 
changes in the momentum-space topology of the single-particle excitation 
spectrum.

\subsection{Topological phase transitions}
\label{sec:top}

As elaborated above, although gap closings occur only at a few $\mathbf{k_0}$ 
points in the lowest quasiparticle and highest quasihole bands,
these gapless excitation points are sufficient to induce intricate polarization 
textures in real space, at and around the boundary between phases with 
locally different momentum-space topology. We emphasize that since the 
symmetry of the order parameters of the local phases have the same 
$s$-wave symmetry across the boundary, the transition is topological.
Such topological changes are known as Lifshitz-type transition in the literature, 
and they are extensively discussed in the context of nodal, e.g. $p$-wave, 
superfluids and superconductors~\cite{volovik}. 
In thermodynamic systems, while the primary signatures of Lifshitz transitions are 
seen in the momentum distribution and single-particle spectral function, some 
thermodynamic quantities, e.g. atomic compressibility and spin susceptibility, also 
show anomalies at the transition boundary. 

It is also worth noting that both spinless $p_x \pm ip_y$ (chiral) 
superfluids~\cite{volovik} and spin-orbit coupled Fermi gases under Zeeman 
field~\cite{iskin, liu} can host Majorana bound states near the edges of the system. 
More specifically, these bound states can only exist at the phase boundary 
between a topologically non-trivial and a trivial phase, the classification 
of which is based on the value of the topological charges, i.e. Chern 
numbers~\cite{volovik}. In contrast to the spinless chiral superfluids, 
our numerical results suggest that the induced transverse polarizations
can be used as a probe to characterize the topological phase transitions 
and the associated Majorana bound states in finite spin-orbit coupled Fermi gases. 
This is in accordance with a recent work on one-dimensional quantum wires 
with strong Rashba and Dresselhaus SOC, where it is shown that the Majorana
polarization can be used as an order parameter to characterize the topological 
transition between the trivial system and the system exhibiting Majorana 
bound modes~\cite{simon}. 

We also remark that spin textures may also occur in $p$-wave superfluids (as
well as in all other systems with vector order parameters) as topological defects, 
i.e. a coreless vortex may exist as a spin texture. For instance, such textures 
were recently observed in superfluid $^3$He~\cite{he-skyrm}, in good agreement
with the early predictions~\cite{merminho, at}. 
Furthermore, in the condensed-matter literature, these two-dimensional 
topological defects were characterized depending on how the local spin 
changes from the center of the defect to its boundary.
Assuming that the local spin $\mathbf{p_i} = p_i \mathbf{\widehat{z}}$ is 
perpendicular to the system at the center of the defect, the topological object 
is called (a) an Anderson-Toulouse spin texture or a baby skyrmion if the spin continuously 
rotates through an angle $\pi$ towards the boundary and anti-aligns with 
respect to the center~\cite{at}, or 
(b) a Mermin-Ho spin texture (meron) or a half-skyrmion if the spin continuously 
rotates through an angle $\pi/2$ and aligns with the plane of the system~\cite{merminho}. 
Note that the polarization textures presented in this work do not belong to these 
classes, and are unique to trapped Fermi gases with SOC.

Having discussed the low-filling Fermi gases at zero temperature, next we
briefly comment on the effects of finite $T$ and high fillings. 
First, although these topological transitions are quantum in their nature, 
signatures of them can still be observed at finite $T$, where the observables 
are smeared out due to thermal effects. 
In particular, we find for the parameters of Figs.~\ref{fig:out} 
and~\ref{fig:in} that the maximum magnitudes of the transverse polarizations 
reduce, respectively, to $50\%$ and $10\%$ at $T = 0.1t$ and $T = 0.2t$.
Second, due to the particle-hole symmetry of the parent Hamiltonian around 
half filling, in addition to the ring-shaped transverse polarizations 
induced near the edges of the system, additional ring-shaped structures 
are further induced near the center of the trap when the center is close to 
a band insulator. Therefore, the transverse polarizations show 
multiple ring-shaped structures in high-filling lattice systems.
Having discussed the numerical results, next we conclude the paper with a
brief summary of our main findings.

\section{Conclusions}
\label{sec:conclusions}

In this paper, we studied the interplay between the Zeeman field, SOC, 
FFLO pairing and polarization textures in harmonically trapped 
two-dimensional Fermi gases, loaded into a single-band tight-binding 
optical lattice. The trapping potential, SOC and Zeeman field are 
taken self-consistently into account via the real-space mean-field BdG 
theory, and two of our main findings can be summarized as follows. 

First, we showed that while the ground states of the spin-orbit coupled 
systems in general have weak FF-type non-uniform superfluid characters 
but not an LO one under the out-of-plane Zeeman field, the FF character 
of the superfluids is stronger for the Rashba SOC under in-plane Zeeman field. 
The FF-type phase oscillations are also most prominent along the direction 
that is perpendicular to the Zeeman field. Therefore, our self-consistent 
results on a finite lattice support recent findings on the thermodynamic 
continuum systems that are ansatz-based non-self-consistent momentum-space 
calculations~\cite{zheng, vjshenoy2b, dong2b, iskinZ, zhangyi, huZ, yiZ,puZ,seoZ}.
We also discussed the effects of interaction, temperature, SOC anisotropy 
and Zeeman field anisotropy on the FFLO-like pairing and ground state 
of the system, and noted that since the superfluid order parameters 
modulate only towards the edges, it may be difficult to detect these modulations
in atomic systems at finite $T$. 

Second, in sharp contrast to the no-SOC case where only an easy-axis polarization 
is possible beyond a threshold Zeeman field, we showed that any non-zero 
combination of the Zeeman field and Rashba SOC induces not only an 
easy-axis polarization everywhere in the system but also a spatially-modulated 
transverse one near the edges. We found that the induced polarization textures 
are skyrmion-like finite-size effects, which are very similar to the spin textures 
that were previously predicted for rotating spinor BEC~\cite{khawaja, mueller} 
and spin-orbit coupled BEC without rotation~\cite{sinha, wu, rama}. 
We also argued that the transverse 
polarizations can be used to probe and characterize the topological phase 
transitions and the associated Majorana bound states in finite spin-orbit 
coupled Fermi gases, and briefly discussed the possibility of observing 
these effects in atomic systems. 

Finally, we emphasize that while all of these results are obtained using an 
optical lattice model, they are equally applicable to continuum systems 
in the low-filling limit. We preferred the lattice description mainly because 
of its easier numerical implementation and versatility, 
e.g. self-consistent inclusion of the trapping potential, and anisotropic 
SOC and/or Zeeman field do not require any additional cost in numerics.
However, since the lattice model allows particle fillings up to unity, 
but with a particle-hole symmetry around half filling, it leads to richer 
finite-size effects compared to the continuum model away from the 
low-filling limit.

\section{Acknowledgments}
\label{sec:ack}
This work is supported by the Marie Curie IRG Grant No. FP7-PEOPLE-IRG-2010-268239, 
T\"{U}B$\dot{\mathrm{I}}$TAK Career Grant No. 3501-110T839, and 
T\"{U}BA-GEB$\dot{\mathrm{I}}$P. 
The author thanks E. Doko, V. B. Shenoy, and A. L. Suba{\c s}{\i} for discussions.


\begin{thebibliography}{99}

\bibitem{nistsoc} Y.-J. Lin,     Y.-J. Lin, K. Jim\'enez-Garc\'ia, and I. B. Spielman, Nature (London) \textbf{471}, 83 (2011).
\bibitem{chinasocb} S. Chen, J.-Y. Zhang, S.-C. Ji, Z. Chen, L. Zhang, Z.-D. Du, Y. Deng, H. Zhai, and J.-W. Pan, Phys. Rev. Lett. \textbf{109}, 115301 (2012). 
\bibitem{chinasocf} P. Wang, Z.-Q. Yu, Z. Fu, J. Miao, L. Huang, S. Chai, H. Zhai, and J. Zhang, Phys. Rev. Lett. \textbf{109}, 095301 (2012).
\bibitem{mitsoc} L. W. Cheuk, A. T. Sommer, Z. Hadzibabic, T. Yefsah, W. S. Bakr, and M. W. Zwierlein, Phys. Rev. Lett. \textbf{109}, 095302 (2012).
\bibitem{engels}  C. Qu, C. Hamner, M. Gong, C. Zhang, and P. Engels, arXiv:1301.0658 (2013).
\bibitem{fu} Z. Fu, L. Huang, Z. Meng, P. Wang, X.-J. Liu, H. Pu, H. Hu, and J. Zhang, arXiv:1303.2212 (2013).
\bibitem{williams} R. A. Williams, M. C. Beeler, L. J. LeBlanc, K. Jim\'enez-Garc\'ia, and I. B. Spielman, arXiv:1306.1965 (2013).

\bibitem{zhaireview} H. Zhai, Int. J. Mod. Phys. B \textbf{26}, 1230001 (2012).
\bibitem{galitskireview} V. Galitski and I. B. Spielman, Nature \textbf{494}, 49 (2013).


\bibitem{volovik} G. Volovik, \textit{The universe in a helium droplet}, Oxford (2003).
\bibitem{hasan} M. Z. Hasan and C. L. Kane, Rev. Mod. Phys. \textbf{82}, 3045 (2010).
\bibitem{sczhang} X.-L. Qi and S.-C. Zhang, Rev. Mod. Phys. \textbf{83}, 1057 (2011).
\bibitem{wen}  X.-G. Wen, arXiv:1210.1281 (2012).

\bibitem{tqc} C. Nayak, S. H. Simon, A. Stern, M. Freedman, and S. das Sarma, Rev. Mod. Phys. \textbf{80}, 1083 (2008).

\bibitem{iskin} M. Iskin, Phys. Rev. A. \textbf{85}, 013622 (2012); 
and Phys. Rev. A \textbf{86}, 065601 (2012).
\bibitem{liu} X.-J. Liu, L. Jiang, H. Pu, and H. Hu, Phys. Rev. A \textbf{85}, 021603(R)(2012);  
X.-J. Liu and H. Hu, Phys. Rev. A \textbf{85}, 033622 (2012).

\bibitem{zheng} Z. Zheng, M. Gong, X. Zou, C. Zhang, and G.-C. Guo, Phys. Rev. A \textbf{87}, 031602(R) (2013).
\bibitem{vjshenoy2b} V. B. Shenoy, arXiv:1211.1831 (2012).
\bibitem{dong2b} L. Dong, L. Jiang, H. Hu, and H. Pu, Phys. Rev. A \textbf{87}, 043616 (2013).
\bibitem{iskinZ} M. Iskin and A. L. Suba\c{s}\i, arXiv:1211.4020 (2012) (to appear in PRA).
\bibitem{zhangyi}  F. Wu, G.-C. Guo, W. Zhang, and W. Yi,  Phys. Rev. Lett. \textbf{110}, 110401 (2013).
\bibitem{huZ} X.-J. Liu and H. Hu, arXiv:1302.0553 (2013).
\bibitem{yiZ}  X.-F. Zhou, G.-C. Guo, W. Zhang, and W. Yi, Phys. Rev. A \textbf{87}, 063606 (2013).
\bibitem{puZ} L. Dong, L. Jiang, and H. Pu, arXiv:1302.1189 (2013).
\bibitem{seoZ} K. Seo, L. Han, and C. A. R. S\'a de Melo, arXiv:1301.1353 (2013).
 
\bibitem{wmliu} X.-L. Yu, S.-S. Zhang, and W.-M. Liu, Phys. Rev. A \textbf{87}, 043633 (2013).
\bibitem{ashrafi} A. Ashrafi, E. I. Rashba, and D. L. Maslov,  arXiv:1306.1165.

\bibitem{FF} P. Fulde and R. A. Ferrell, Phys. Rev.\textbf{135}, A550 (1964).
\bibitem{LO} A. I. Larkin and Y. N. Ovchinnikov, Zh. Eksp. Teor. Fiz. \textbf{47}, 1136 (1964); 
and Sov. Phys. JETP \textbf{20}, 762 (1965).

\bibitem{kohl} M. Feld, B. Frohlich, E. Vogt, M. Koschorreck, and M. K\"ohl, Nature \textbf{480}, 75 (2011).
\bibitem{vogt} E. Vogt, M. Feld, B. Frohlich, D. Pertot, M. Koschorreck, and M. K\"ohl; Phys. Rev. Lett. \textbf{108}, 070404 (2012).
\bibitem{sommer} A. T. Sommer, L. W. Cheuk, M. J.-H. Ku, W. S. Bakr, and M. W. Zwierlein, Phys. Rev. Lett. \textbf{108}, 045302 (2012).

\bibitem{skyrme} T. H. R. Skyrme, Proc. R. Soc. A 260, \textbf{127} (1961); 
and Nucl. Phys. \textbf{31}, 556 (1962).

\bibitem{sondhi} S. L. Sondhi, A. Karlhede, S. A. Kivelson, and E. H. Rezayi, Phys. Rev. B \textbf{47}, 16419 (1993).
\bibitem{barrett} S. E. Barrett, G. Dabbagh, L. N. Pfeiffer, K. W. West, and R. Tycko, Phys. Rev. Lett. \textbf{74}, 5112 (1995). 

\bibitem{khawaja} U. Al Khawaja and H. T. C. Stoof, Phys. Rev. A \textbf{64}, 043612 (2001).
\bibitem{mueller} E. J. Mueller, Phys. Rev. A \textbf{69}, 033606 (2004).
\bibitem{sinha}  S. Sinha, R. Nath, and L. Santos, Phys. Rev. Lett. \textbf{107}, 270401 (2011).
\bibitem{wu} C. Wu, I. Mondragon-Shem, and X.-F. Zhou, Chin. Phys. Lett. \textbf{28}, 097102 (2011).
\bibitem{rama} B. Ramachandhran, B. Opanchuk, X.-J. Liu, H. Pu, P. D. Drummond, and H. Hu,  Phys. Rev. A \textbf{85}, 023606 (2012). 


\bibitem{edoko} E. Doko, A. L. Suba{\c s}{\i}, and M. Iskin, Phys. Rev. A \textbf{85}, 053634 (2012). 

\bibitem{simon} D. Sticlet, C. Benna, and P. Simon, Phys. Rev. Lett. \textbf{108}, 096802 (2012).


\bibitem{he-skyrm} R. Blaauwgeers, V. B. Eltsov, M. Krusius, J. J. Ruohio, R. Schanen, and G. E. Volovik, Nature \textbf{404}, 471 (2000).
\bibitem{merminho} N. D. Mermin and T. L. Ho, Phys. Rev. Lett. \textbf{36}, 594 (1976). 
\bibitem{at} P. W. Anderson and G. Toulouse, Phys. Rev. Lett. \textbf{38}, 508 (1977).

\end{thebibliography}
\end{document}